\documentclass[aps,prc,reprint,groupedaddress]{revtex4-2}
\newcommand*{\rom}[1]{\expandafter\@slowromancap\romannumeral #1@}
\makeatother
\usepackage{amsmath}
\usepackage{graphicx}
\usepackage{dcolumn}
\usepackage{bm}
\usepackage{algpseudocode}
\usepackage[caption=false]{subfig}
\usepackage{url}
\usepackage{cases}
\usepackage{longtable}
\usepackage{float}
\usepackage{booktabs}

\begin{document}

\title{Interpretable hybrid nuclear mass prediction based on term-by-term model discrepancies}

\author{Weihu Ye}
\affiliation{School of Physics and Optoelectronics, South China University of Technology, Guangzhou 510641, China}
\author{Niu Wan}
\email{wanniu@scut.edu.cn}
\affiliation{School of Physics and Optoelectronics, South China University of Technology, Guangzhou 510641, China}
\newcommand{\RNum}[1]{\uppercase\expandafter{\romannumeral #1\relax}}
\date{\today}

\begin{abstract}
Various theoretical mass models have consistently achieved impressive accuracy in reproducing experimental masses. However, their predictions in unmeasured neutron-rich regions exhibit noticeable model dependence. In this study, we systematically investigate the differences in model predictions by comparing the liquid-drop mass terms of two representative models. Using two widely‐used models, Weizs$\ddot{a}$cker–Skyrme-type (WS4) and the Duflo–Zuker-type (DZ10) as representative examples, we find that the differences in an isotope chain gradually become more remarkable with increasing neutron number not only for total binding energies but also for individual mass terms, and the most noticeable term is the volume-symmetry energy. By leveraging the term-by-term differences between the liquid-drop energy components of WS4 and DZ10, we introduce a machine learning gating network that adaptively combines the two models to improve predictive accuracy. This conditional hybrid model achieves a lower root-mean-square deviation (rmsd) than either model alone, reducing the overall rmsd from 0.284 MeV (WS4) and 0.560 MeV (DZ10) to 0.232 MeV. In the future, this term-by-term comparison strategy can be extended to models based on density-functional theories.
\end{abstract}
\maketitle
%
%
%
%

%

\section{INTRODUCTION}
Nuclear mass plays a paramount role in both nuclear physics \cite{Lunney,zhang2024predictions,zhang2024possibility} and astrophysics \cite{kajino2019current,cowan2021origin}. It serves as an effective tool to provide a variety of nuclear structure information, e.g., shell effect \cite{otsuka2020evolution}, deformation \cite{moller2016nuclear}, residual interaction \cite{Garvey}, and so on. Furthermore, nuclear mass itself is an essential factor in simulations of rapid neutron capture process \cite{brett2012sensitivity,mumpower2015impact}, and the related $Q$ values such as nucleon separation energies are also crucial in the process \cite{mumpower2012influence,lund2023influence}, which is responsible for understanding the origin of elements heavier than iron in the Universe \cite{mumpower2016impact,meyer1994r}.

A wide variety of theoretical approaches has been developed to reproduce nuclear masses.
Macroscopic mass models, such as the liquid-drop model, rely solely on bulk nuclear properties and generally reproduce experimental masses with root-mean-square deviations (rmsds) of about 1-3 MeV \cite{bethe1936nuclear,kirson2008mutual}. Macroscopic-microscopic (mac-mic) models further incorporate shell and pairing corrections, reducing the rmsds to about 0.3-0.5 MeV. Representative examples include the finite-range droplet model \cite{moller2016nuclear}, the Weizsäcker-Skyrme (WS) series \cite{wang2014surface,MinWang,wang2010modification,wang2010mirror}, and the Duflo-Zuker (DZ) family \cite{duflo1995microscopic}. Microscopic models, particularly those based on covariant density functional theory (CDFT) \cite{guo2024nuclear,zhang2014global,liang2013feasibility,goriely2013further,ryssens2022skyrme,yang2021nuclear}, have also achieved remarkable progress in recent years. By treating nuclear interactions through relativistic energy-density functionals derived from meson-exchange mechanisms, CDFT naturally incorporates relativistic dynamics that are essential for describing heavy and exotic nuclei \cite{meng2015halos,agbemava2015covariant,afanasjev2013nuclear}. In addition to these global approaches, local mass relations \cite{Garvey,Barea,bao2016simple,bao2013,fu2010nuclear} provide an alternative strategy, predicting masses through algebraic connections among neighboring nuclei. In addition to these physics-based models, data-driven machine learning methods have recently emerged as a crucial tool in mass studies, offering flexible frameworks for improving predictive accuracy \cite{wang2023machine,niu2019comparative,he2023machine,boehnlein2022colloquium,gao2021machine,niu2022nuclear,Piekarewicz,sharma2022learning,li2022deep}. In this context, ML has been applied mainly in two directions. One direction is to directly learn measured masses using ML models, such as support vector machines, Gaussian process regression, kernel ridge regression, and neural networks. The other direction is to learn the residuals between theoretical predictions and experimental masses, so that the ML model can primarily capture systematic trends not fully captured by the underlying model. In addition, uncertainty estimation for extrapolations has attracted increasing attention, with approaches such as mixture-density networks and Bayesian strategies being explored to provide calibrated predictive uncertainties. In general, these ML approaches can reach accuracies at the level of 200 keV for measured nuclei.

One can see that the rmsds of the above mass models have decreased significantly with the steady progress in nuclear theory, making it possible to predict unmeasured masses well for applications, such as the synthesis of superheavy elements \cite{oganessian2010synthesis} and the rapid neutron capture process. Nevertheless, these mass predictions remain highly uncertain because the models incorporate different physical mechanisms and energy components \cite{sobiczewski2014predictive,sobiczewski2014accuracy,ye2022accuracy}. Specifically, mass differences between models can steadily increase from less than one MeV to several dozen MeV when extrapolating towards the neutron drip line \cite{goriely2007further,goriely2003further,goriely2010further,goriely2013further,MinWang}. Hence, assessing model discrepancies has become increasingly important, particularly for unmeasured masses. Recent efforts have employed machine learning techniques~\cite{he2023machine,li2022deep,gao2021machine,Niu,li2025investigation,niu2018high,Niu2019,huang2025validation,li2022deep} to improve predictive reliability. 

While recent machine-learning approaches have improved predictive accuracy, most studies focus either on direct mass learning from data or on residual corrections relative to a single baseline model. Here, rather than modifying a single model, we improve predictive accuracy by integrating the complementary physical ingredients of different mass models. In this work, we firstly perform a systematic term-by-term comparison between two representative global mass models, WS4~\cite{wang2014surface} and DZ10~\cite{duflo1995microscopic}, to identify the dominant sources of their discrepancies. Guided by these physical insights, we introduce a machine-learning gating network that adaptively combines the WS4 and DZ10 predictions. The gating network takes the differences between their liquid-drop components as input features and outputs a nucleus-specific weight, allowing the model to select the more reliable description in different regions of the nuclear chart. Compared with purely data-driven approaches, this framework retains the structure of the original mass models and provides a clear map of model preference across the nuclear chart. The resulting hybrid achieves a lower rmsd than either individual model and offers insight into the regional reliability of competing theoretical descriptions.

This paper is structured as follows. Section~II provides the necessary descriptions of the employed mass models. Section~III presents the detailed term-by-term comparison between WS4 and DZ10, followed by the construction of a machine learning gating network that adaptively combines the two mass models using the residual liquid-drop features. Finally, conclusions are given in the last section.

\section{DESCRIPTION OF THEORETICAL MODELS}
\subsection*{A. Weizs$\ddot{a}$cker-Skyrme mass model}
The WS4 \cite{wang2014surface} is a mac-mic model with the rmsd between theoretical and experimental masses of approximately 0.3 MeV. It has been widely applied in nuclear astrophysical simulations \cite{zhao2019r}, mass-extrapolation tests \cite{li2025investigation}, and the search for long-lived superheavy nuclei \cite{wang2016correlations}. The WS4 incorporates a modified liquid-drop (LD) model with deformation corrections and accounts for shell effects based on the Strutinsky method \cite{brack1972funny}. This combination allows for a comprehensive description of nuclear binding energies (BE). Specifically, the LD energy can be interpreted as the smooth, linear variation of binding energies with increasing nucleon numbers, whereas the shell energy represents the irregular fluctuations at specific nucleon numbers. The total binding energy of a nucleus is calculated by summing the LD energy and the shell corrections, as formulated below \cite{wang2010modification,wang2010mirror},
\begin{align}
BE(N,Z,\beta)
  &= E_{LD}(N,Z)\prod_{k\ge 2} \left(1 + b_{k}\beta_{k}^{2}\right)  \nonumber\\
  &\quad + \Delta E_{\mathrm{shell}}(N,Z,\beta),
\label{00}
\end{align}
where $N$ and $Z$ denote the neutron and proton numbers, respectively. In contrast to spherical nuclei, the coefficients $b_k$ and $\beta_k$ account for the additional contributions arising from nuclear deformation, which are determined using the Skyrme energy-density functional \cite{moller2016nuclear}. The explicit expression for the LD energy is given by
\begin{align}
E_{LD}
  &= a_1 E_{vol} - a_2 E_{sur}
     - a_3 E_{Coul} - a_4 E_{sym} \nonumber\\
  &\quad + a_5 E_{ssym} - a_6 E_{pair} \nonumber\\
  &= a_1 A - a_2 A^{2/3}
     - a_3 \frac{Z^2}{A^{1/3}}\!\left(1 - 0.76\,Z^{-2/3}\right) \nonumber\\
  &\quad - a_4 I^2 A \!\left[1 + \frac{\xi(2 - |I|)}{2 + |I|A}\right] f_s
     + a_5 I^2 A^{2/3} f_s \nonumber\\
  &\quad - a_6 A^{-1/3} \delta_{np},
\label{01}
\end{align}
with
\begin{align}
f_s &= 1 + \kappa_s\,\varepsilon\,A^{1/3}, 
\qquad \varepsilon = (I - I_0)^2 - I^4, \nonumber\\[2pt]
I   &= \frac{N - Z}{A}, 
\qquad I_0 = \frac{0.4A}{A + 200},
\label{eq:fs}
\end{align}
where the energy terms on the right side of Eq.~\ref{01} correspond to the volume, surface, Coulomb, volume-symmetry, surface-symmetry, and pairing terms, respectively. The term $\delta_{np}$ in the pairing energy accounts for the odd-even properties of proton and neutron numbers. 

\begin{table}[ht]
\centering
\caption{Parameters for the WS4 and DZ10 models.}
\begin{tabular*}{\columnwidth}{@{\extracolsep{\fill}}lcc}
\hline
Parameter & WS4 \cite{wang2014surface} & DZ10 \cite{ye2022accuracy} \\ 
\hline
$a_1$      & 15.5181 & 17.7479 \\ 
$a_2$      & 17.4090 & 16.2513 \\ 
$a_3$      & 0.7092  & 0.7054  \\ 
$a_4$      & 30.1594 & 37.3563 \\ 
$a_5$      & 45.8091 & 52.6613 \\ 
$a_6$      & 5.8166  & 6.1026  \\ 
$\kappa_s$ & 0.1536  &         \\ 
$\xi$      & 1.2230  &         \\ 
$\kappa$   & 1.5189  &         \\ 
\hline
\end{tabular*}
\label{para}
\end{table}

\subsection*{B. Duflo-Zuker mass model}
The DZ10 \cite{duflo1995microscopic,kirson2012empirical} is a mac-mic mass model with an rmsd of $\approx$ 0.56 MeV. It is constructed based on the assumptions of the shell model. Utilizing a predefined shell structure, valence nucleons are sequentially filled into harmonic oscillator shells, providing insight into shell effects in a parameterized way. The binding energy of DZ10 can be divided into two categories, one is related to the LD model, and the other is associated with the interacting shell model. These two groups are labeled as $E_{LD}$ and $<H_{m}>$, and are formulated as follows:
\begin{align}
BE &= E_{LD} + < H_m >,
\label{4}
\end{align}
with
\begin{align}
E_{LD}
  &= a_{1} E_{vol} + a_{2} E_{sur}
     + a_{3} E_{Coul} - a_{4} E_{sym} \nonumber\\
  &\quad + a_{5} E_{ssym} + a_{6} E_{pair} \nonumber\\
  &= a_{1}(M+S) - a_{2} M/\rho + a_{3}\big(-Z(Z-1) \nonumber\\
  &\quad  + 0.76[Z(Z-1)]^{2/3}\big)/\rho - a_{4} T(T+2)/A^{2/3} \nonumber\\
  &\quad  + a_{5} T(T+2)/A^{2/3} \rho^{2}  + a_{6}\,\delta_{np}/\rho,
\label{5}
\end{align}
and
\begin{align}
< H_m>
  &= a_7 s_3 - a_8\, s_3/\rho + a_9 s_4 + a_{10} d_4,
\label{6}
\end{align}
where $\rho= A^{1/3}[1-0.25\frac{T^2}{A^2}]^2$ denotes a scaling factor with the isospin term $T=|N-Z|$. In the $E_{\rm LD}$ group, the terms $E_{\rm vol}$ and $E_{\rm sur}$ represent the macroscopic volume and surface contributions (expressed through the master-term combinations $M+S$ and $M/\rho$), while $E_{\rm Coul}$, $E_{\rm sym}$, $E_{\rm ssym}$, and $E_{\rm pair}$ denote the Coulomb, volume-symmetry, surface-symmetry, and pairing energies, respectively. The symbol $\delta_{np}$ in the pairing energy is determined by the odd–even property of proton and neutron numbers. In the $<H_m>$ group,  the $s_3$, $\frac{s_3}{\rho}$ and $s_4$ originate from the spherical consideration, while the last term $d_4$ describes the deformation effect. The coefficients $a_i$ are free parameters fitted to experimental mass data, and their values for WS4 and DZ10 are listed in Table~\ref{para}.

\begin{figure*}
\includegraphics[width=17cm]{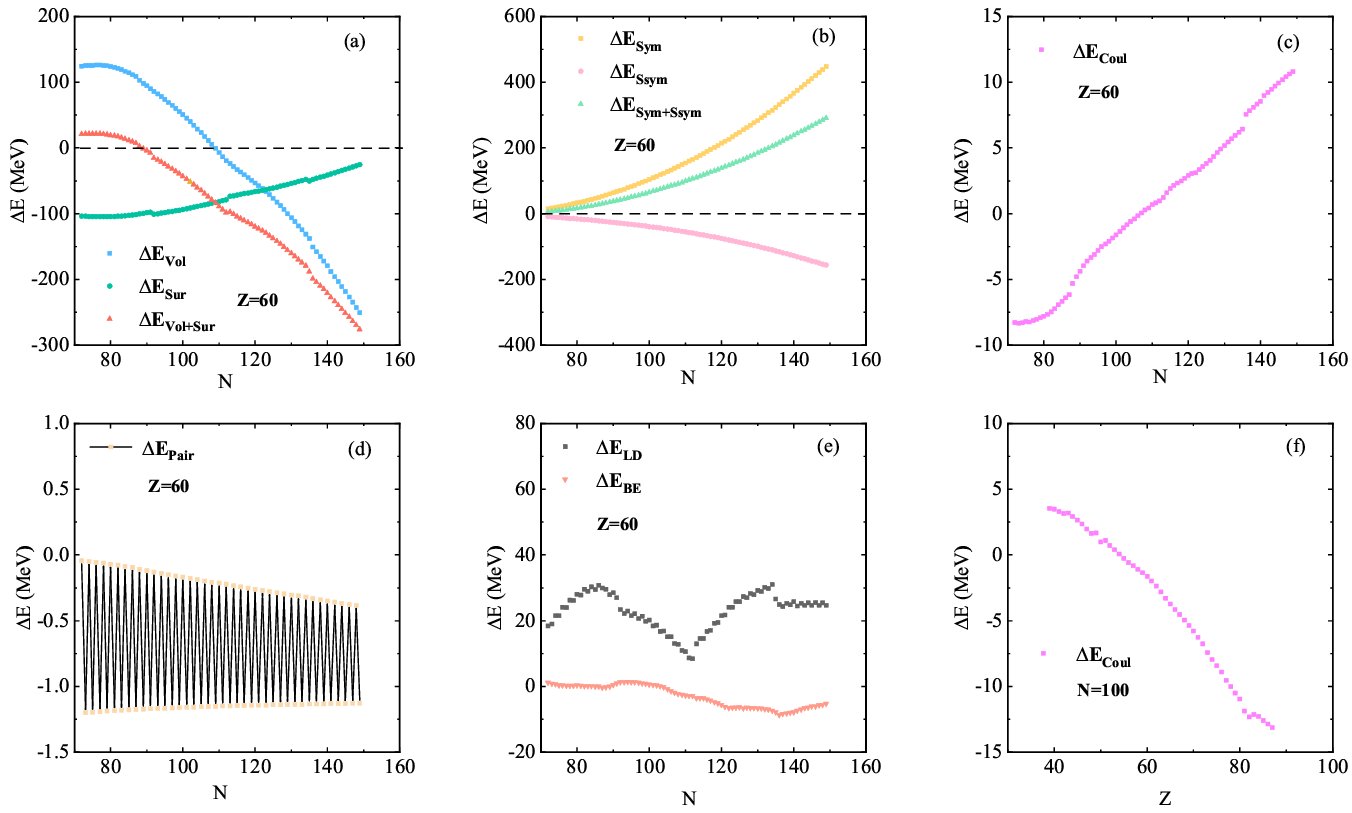}
\centering
\caption{\label{fig1} Evolution of the term differences (in MeV) between the WS4 and DZ10 models for the $Z$=60 isotope chain. Panels (a–e) show the differences in the volume, surface, volume-symmetry, surface-symmetry, Coulomb, and pairing terms (defined as $\Delta E_{ {term}} = E_{ {WS4,term}} - E_{ {DZ10,term}}$), as well as their sum, denoted as vol, sur, sym, ssym, Coul, pair, and LD, respectively. For comparison, the differences in the total BE are also displayed. Panel (f) presents the differences in the Coulomb term for the isotone chain with $N=100$.}
\end{figure*}
\section{RESULTS AND DISCUSSIONS}
\subsection{Term-by-term comparison}
\label{Term-by-term}
As previously mentioned, this work aims to provide deeper insights into prediction differences by directly comparing the individual mass terms between different models. Given that some terms differ considerably between models, it is not feasible to fully compare all terms. For instance, while the WS4 model uses a single term for shell corrections, the DZ10 model incorporates a series of terms for shell considerations. Consequently, a selected subset of terms, referred to as the LD part, is adopted, including the volume, surface, volume-symmetry, surface-symmetry, Coulomb, and pairing terms. The subsequent calculations will primarily focus on these six terms.

We begin by calculating the differences in the six terms between the models. Taking the comparison of the volume term as an example, the difference for a nucleus is extracted from the subtraction between the volume term in the WS4 and the corresponding volume terms in the DZ10, namely, $\Delta E_{vol}=E_ {{vol}}^{WS4}-E_ {{vol}}^{DZ10}$. The remaining five terms are treated in the same way. For nuclei along the isotope chain of Neodymium (Nd, $Z$=60), this calculation is performed. Fig.~\ref{fig1} illustrates the evolution of these differences with neutron number for the six terms.

The experimentally determined limit for Nd isotopes approaching the neutron drip line currently stands at $N=100$. We begin by comparing the volume and surface terms. As shown in Fig.~\ref{fig1}(a), the differences in the volume terms decrease steadily to 0 MeV before increasing in the negative direction. The differences in the surface terms are initially around -100 MeV and gradually decrease to approximately -50 MeV as the neutron number increases. These suggest that the magnitudes of the volume and surface terms in the DZ10 model differ from those in the WS4 model, both in the measured and unmeasured regions. Considering that both volume and surface terms share the same physical origin, their combined difference is also investigated. When the differences in volume and surface terms are summed, their combined deviations are always 0 MeV for the measured nuclei with $N<100$, and they increase with the neutron number. Similarly, the volume-symmetry and surface-symmetry terms are handled as well. As shown in Fig.~\ref{fig1}(b), the differences in the volume-symmetry terms increase drastically as the neutron number grows, with magnitudes becoming noticeable at the neutron-drip line, reaching up to approximately 500 MeV. In contrast to the volume and surface terms, the trends in volume-symmetry energy and its surface correction are globally opposite, thus narrowing down their combined difference to some extent.
 \begin{figure*}
 \centering
     \includegraphics[width=14cm]{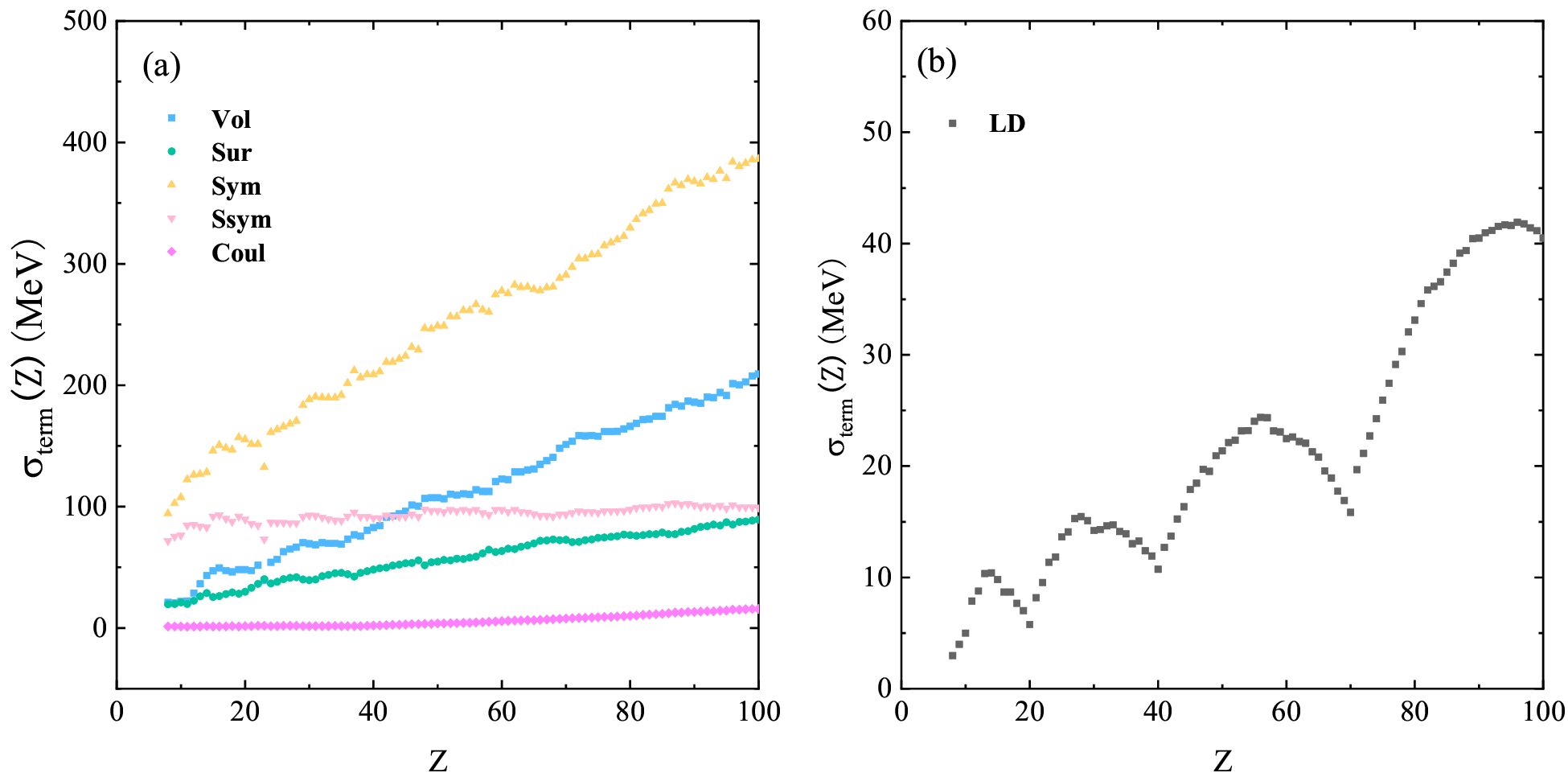}
     \caption{\label{fig3} Rmsds (in MeV) of the term differences, calculated according to Eq.~(\ref{9}), for volume, surface, volume-symmetry, surface-symmetry, Coulomb terms, and LD energies between the WS4 and DZ10 models for isotope chains ranging from $Z$=8 to $Z$=100. }
 \end{figure*}

The Coulomb energy is usually considered as one of the best constrained terms in macroscopic mass models. Nevertheless, as shown in Fig.~\ref{fig1}(c), the differences in the Coulomb contribution continue to increase with the neutron number, indicating that even this term is affected by model-dependent extrapolations in neutron-rich regions. As shown in Fig.~\ref{fig1}(d), the differences in the pairing term remain very small, indicating that its uncertainty is much smaller than those of the other macroscopic terms. Also evident is the odd-even staggering seen in the residuals. The overall deviations in the sum of the six terms and the total binding energies are provided in Fig.~\ref{fig1}(e). The trend in the differences of binding energies is moderate, with deviations gradually increasing. In contrast, the trend in the differences of LD energies shows relatively large deviations, particularly around the two peaks near the magic numbers. This suggests that certain terms outside the LD part, such as the shell corrections, do compensate for the large deviations in the LD energies, thereby reducing the overall residuals of the binding energies. From these term-by-term comparisons, one can conclude two issues for WS4. No clear cancellation of terms is observed in either the measured or unmeasured regions. The correlation between the term differences and neutron numbers is found, as illustrated by the fact that the differences for some terms such as the volume-symmetry and surface-symmetry terms steadily increase with rising neutron numbers.

The term differences along the isotope chains were first assessed, and a logical next step is to examine their evolution along the isotone chains. Similar to the previous comparisons, calculations were carried out for several isotone chains. The resulting trends exhibit similar magnitudes but opposite directions to those observed for the isotope chains. For example, Fig.~\ref{fig1}(f) shows the differences in the Coulomb terms for a fixed neutron number $N=100$. The deviations decrease progressively with increasing proton number, in contrast to the trend observed in Fig.~\ref{fig1}(c). This opposite behavior mainly reflects that the isotone chain starts in the neutron-rich region, where most nuclei are unmeasured, and extends toward the $\beta$-stable region with better experimental constraints.

Note that the term differences within the LD part are relatively large. However, it is not possible to definitively determine which physical term in the DZ10 is more reliable than in the WS4, or vice versa. Both models are constructed with high accuracy in mass reproduction and exhibit a certain level of reliability in extrapolation. Therefore, the information derived from the term differences primarily helps to clarify the origin of the model discrepancies rather than to judge which model is superior.
 
Having examined the term differences along representative isotope and isotone chains in Fig.~\ref{fig1}, we now extend the analysis to a systematic investigation over the entire nuclear chart. A new statistical approach is then adopted to quantify the term differences across isotope chains. Specifically, isotope chains with proton numbers ranging from $Z$=8 to $Z$=100 are examined. Beginning with $Z$=8 (O), term differences are calculated for nuclei, ranging from experimentally unmeasured neutron-rich isotopes up to the neutron-drip line. The measured heaviest nucleus for O is $^{30}$O \cite{wang2021ame}, so the unmeasured neutron-rich isotopes of O start with $^{31}$O and end with $^{32}$O which corresponds to the neutron-drip line predicted by the theoretical model. Next, the rmsds of the term differences are calculated to quantify the average variations along each chain, as defined by $\sigma_{\mathrm{term}}(Z)$, 
\begin{eqnarray}
\sigma_{\mathrm{term}}(Z)  = \sqrt{\frac{\sum_{i=1}^{n}\left(E_{i}^{ {WS4}} - E_{i}^{ {DZ10}}\right)^{2}}{n}}.
\label{9}
\end{eqnarray}
The same procedure is repeated for all subsequent chains from $Z$=9 to $Z$=100. This approach enables us to capture the overall evolution of term differences in the unmeasured regions across the isotope chains.

The rmsds for volume, surface, volume-symmetry, surface-symmetry, Coulomb, and the total LD energies are provided in Fig.~\ref{fig3}. The results for the pairing term show minimal variation around 0 MeV and were therefore excluded. First, a clear ordering of terms is observed. The rmsds of the volume-symmetry term are consistently the largest across all isotope chains, indicating that it contributes most strongly to the differences between the two models. Second, at low proton numbers, the rmsds of the volume and surface terms are comparable. However, with increasing proton number, the rmsds of the volume term exceed those of the surface term. The rmsds of the surface-symmetry term increase moderately, exceeding the volume and surface terms at lower $Z$ but falling below them at higher $Z$. These trends suggest that the relative importance of the LD terms varies across different isotope chains. Third, a gradual increase in the rmsds of all terms is observed, indicating that the error in the term differences increases along the isotope chains. This trend is also reflected in the Coulomb term, whose rmsds increase slowly with $Z$. For the total LD energy, the most pronounced deviations, four peaks as shown in Fig.~\ref{fig3}(b), appear around nuclei close to the magic numbers, indicating enhanced theoretical uncertainties associated with shell effects.
\subsection{Gating network hybrid based on term residuals}
Following the systematic comparison of the six macroscopic LD terms between WS4 and DZ10 in Section~\ref{Term-by-term}, we now explore how these differences can be leveraged to improve model accuracy. To this end, we introduce a machine learning gating network \cite{jacobs1991adaptive,shazeer2017outrageously} that adaptively combines the WS4 and DZ10 predictions across the nuclear chart. Gating networks, originally developed within the mixture-of-experts framework, assign data-driven weights to multiple predictors based on the input features. In this study, the term differences (hereafter referred to as the $\Delta$-terms) are employed as input features, allowing the network to learn which model should be given greater weight in the combined prediction.

The gating network is designed to assign a data-driven weight to the two model predictions for each nucleus. We collect the six liquid-drop residuals $\Delta$-terms ($\Delta E_ {vol}$, $\Delta E_ {sur}$, $\Delta E_ {sym}$, $\Delta E_ {ssym}$, $\Delta E_ {Coul}$, and $\Delta E_ {pair}$), each representing the difference between the WS4 and DZ10 values for the corresponding macroscopic liquid-drop term. These $\Delta$-terms form the input feature vector $\mathbf{x}$, and the blended binding energy prediction is written as
\begin{equation}
BE_{blend} = w( {x})\, BE_{DZ10} + [1 - w( {x})]\, BE_{WS4},
\end{equation}
where \(w(\mathbf{x}) \in [0,1]\) is a function of the \(\Delta\)-term vector \(\mathbf{x}\) for the given nucleus. 

In practice, the network consists of three fully connected layers with 64, 32, and 1 neurons, respectively. Each hidden layer employs the ReLU activation function, and the output layer uses a sigmoid activation to produce the gating weight. The model is trained using the Adam optimizer with a learning rate of $3\times10^{-3}$. Early stopping is applied, and convergence is typically reached within about 1000 iterations.

Although the six liquid-drop residual terms are not physical observables, they quantify how WS4 and DZ10 differ in describing the macroscopic energy components. These $\Delta$-terms therefore serve as diagnostic features that encode the local discrepancy between the two models across the nuclear chart. Using these features, the gating network learns statistical correlations between the residual patterns and the relative performance of WS4 and DZ10. Consequently, it assigns a smaller weight $w(\mathbf{x})$ to DZ10 in regions where WS4 is more reliable, and a larger weight otherwise, thereby identifying the trust regions of the two models in a physically interpretable manner.

\begin{table}[ht]
\centering
\caption{RMSDs (MeV) for WS4, DZ10, the fixed-average baseline, and the gated hybrid model. The values for the hybrid and fixed-average models are reported as the mean and standard deviation over 10-fold cross-validation.}
\label{tab:gate_results_3line}
\begin{tabular}{lccc}
\toprule
Model & $\sigma_{\text{train}}$ (MeV) & $\sigma_{\text{test}}$ (MeV) & $\sigma_{\text{all}}$ (MeV) \\
\midrule
WS4   &  &  & 0.284 \\
DZ10  & &  & 0.562 \\
Case-I & $0.354 \pm 0.001$ & $0.353 \pm 0.017$ & $0.3542 \pm 0.000$ \\
Case-II   & $0.236 \pm 0.015$ & $0.258 \pm 0.015$ & $0.239 \pm 0.014$ \\
\bottomrule
\end{tabular}
\end{table}

\begin{figure*}
\includegraphics[width=10cm]{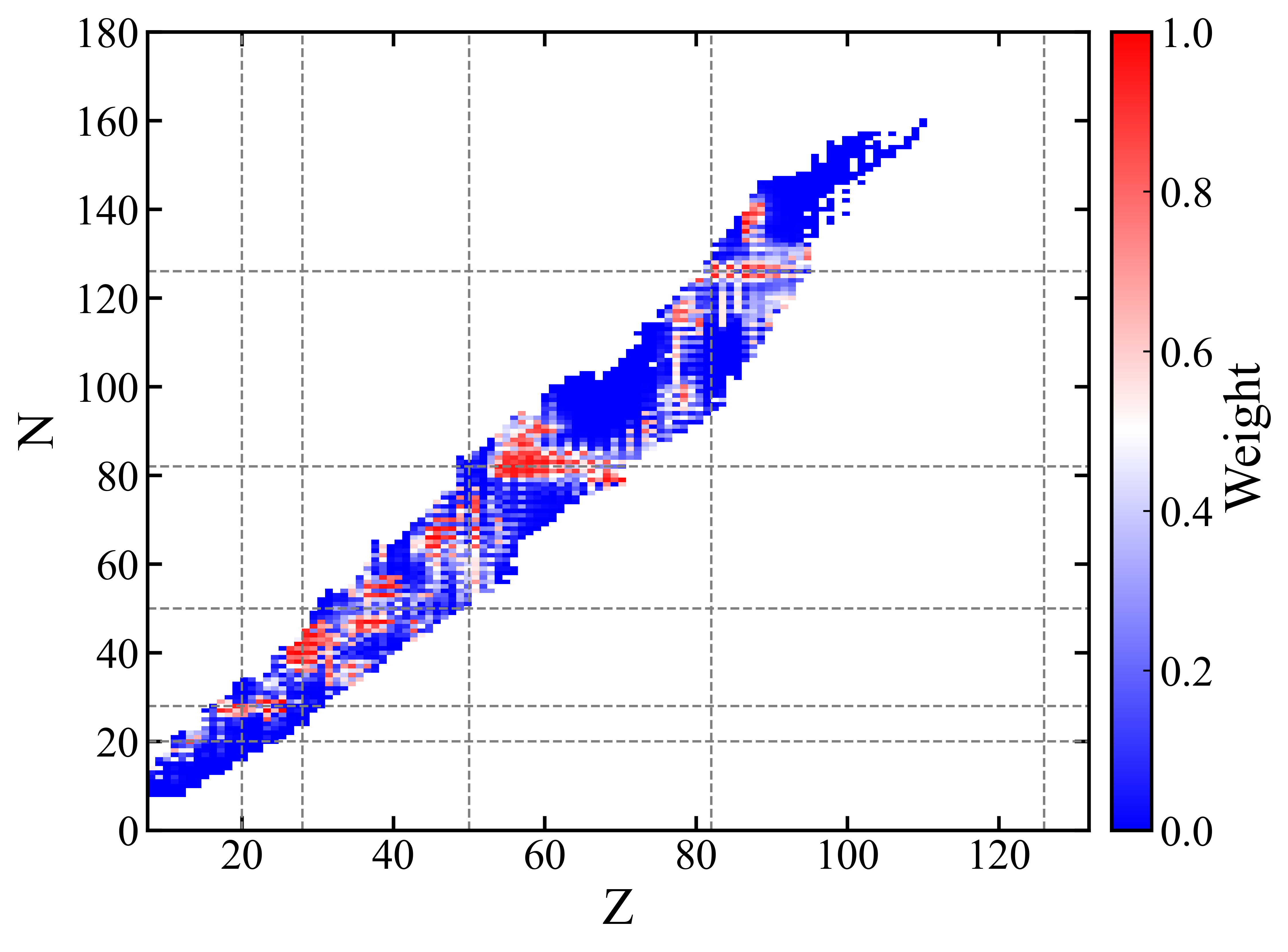}
\centering
\caption{\label{fig33}Gating weight distribution $w(N,Z)$ for the WS4+DZ10 hybrid model using the BE+6$\Delta$-terms feature set. The color scale represents the gating weight assigned to DZ10 ($w$), where red corresponds to regions dominated by DZ10 ($w \rightarrow 1$) and blue to regions dominated by WS4 ($w \rightarrow 0$). The gray guiding lines correspond to the magic numbers 20, 28, 50, 82, and 126. }
\end{figure*}
To assess the effect of including the residual $\Delta$-terms, two feature settings were designed for comparison: one using only the binding energies from WS4 and DZ10 (BE only), and the other additionally incorporating the six liquid-drop $\Delta$-terms (BE+6$\Delta$-terms). A total of 2340 experimental nuclei with $(N,Z \ge 8)$ and uncertainties below 0.1~MeV are selected from the AME2020~\cite{wang2021ame} dataset, of which 80\% were used for training and 20\% for testing. The model accuracy is also measured by the rmsd, defined as
\begin{equation}
\sigma_{BE} = \sqrt{\frac{1}{n}
\sum_{i=1}^{n}\!\left(BE_{i}^{theo} - BE_{i}^{expt}\right)^{2}},
\label{sigma}
\end{equation}
where $BE^{theo}$ and $BE^{expt}$ denote the theoretical and experimental binding energies, respectively.
 
Table~\ref{tab:gate_results_3line} shows the results of the hybrid model together with the fixed-average baseline. To ensure the robustness and generalization of the results, a 10-fold cross-validation was employed. This procedure evaluates the model’s stability and minimizes the risk of overfitting by testing on data subsets independent of the training process. The reported mean and standard deviation reflect the model's consistent performance across 10 random data partitions. The training and testing accuracies of the gated hybrid are comparable, indicating that the gating network achieves stable convergence and good generalization capability. The fixed-average baseline gives an overall accuracy of $\sigma_{\mathrm{all}}=0.354$~MeV, which is even worse than that of WS4 itself (0.284~MeV). This suggests that directly combining the two models without adaptive diagnostic information brings little improvement. In contrast, after incorporating the six liquid-drop residual $\Delta$-terms, the gated hybrid reduces the overall deviation to $0.239 \pm 0.014$~MeV, demonstrating that the macroscopic differences provide valuable information for enhancing the adaptive hybrid performance.

To further enhance the physical interpretability of the machine learning model, we analyzed the spatial distribution of the gating weights ($w$) across the nuclear chart. Fig.~\ref{fig33} shows the gating-weight map obtained from the BE+6$\Delta$-terms setting. The WS4 model (blue regions) dominates over a broad region, including light, medium-mass, and superheavy nuclei, where it provides smoother and more accurate macroscopic trends. In contrast, the DZ10 model (red regions) acquires larger weights in several closed-shell regions, particularly around $N=82$, where the effect is most pronounced. This behavior is consistent with the fact that the DZ10, which is based on a shell model with valence nucleons filling harmonic oscillator shells, is inherently well-suited for describing shell effects. Overall, the predictive behaviors of WS4 and DZ10 are complementary, enabling a more precise and robust hybrid model that improves the predictive accuracy. These results indicate that the additional $\Delta$-term information helps the gating network better capture local macroscopic differences, leading to a more adaptive and physically interpretable combination of the two models.

\section{SUMMARY}
To quantify the prediction differences between models in a more rigorous way, we perform a systematic comparison between the WS4 and DZ10 formulas for the volume, surface, volume-symmetry, surface-symmetry, Coulomb, and pairing terms. The macroscopic liquid-drop components in WS4 and DZ10 are found to differ in magnitude, with the largest discrepancy appearing in the volume-symmetry term, especially for neutron-rich nuclei, while other terms show smaller but non-negligible differences in specific regions. In addition, we analyze the variation trends of all mass terms along isotope chains from $Z$=8 to $Z$=100, and find that the relative importance of these terms also changes from one chain to another.

Building on the above analysis, we introduce a machine learning gating hybrid network that adaptively combines the WS4 and DZ10 models by using the differences between their liquid-drop components as diagnostic features. This approach transforms the differences between the two models into useful signals, allowing the network to give more weight to the model that performs better in each region. Our model achieves a lower overall rmsd of 0.232~MeV, compared with 0.284~MeV for WS4 and 0.560~MeV for DZ10. Moreover, the gating-weight map reveals a physically interpretable pattern: the WS4 dominates in the regions of light, medium, and superheavy nuclei, whereas the DZ10 contributes more near closed-shell regions. These results show that the term-by-term diagnostic comparison not only clarifies the physical origin of model differences but also provides effective features for building interpretable machine learning models.

\begin{acknowledgments}
This work was supported by the National Natural Science Foundation of China (Grant No. 12205105), by the Fundamental Research Funds for the Central Universities (Grant No. 2024ZYGXZR058) and the startup funding of South China University of Technology.
\end{acknowledgments}

\bibliographystyle{apsrev4-2}
\bibliography{ref}
\end{document}